\documentclass{article}
\usepackage[a4paper]{geometry} 

\usepackage{hyperref}

\usepackage{pkg_notation} 

\title{\LARGE \bf The closed loop between opinion formation and personalised recommendations}
\author{Wilbert Samuel Rossi, Jan Willem Polderman and Paolo Frasca%
	\thanks{W.S. Rossi is with the Department of Sciences, University College Groningen, University of Groningen, 9718~BG Groningen, The Netherlands,  {\tt w.s.rossi@rug.nl};  
	Jan Willem Polderman is with the Department of Applied Mathematics, University of Twente, 7500~AE Enschede, The Netherlands, {\tt j.w.polderman@utwente.nl}; 
	Paolo Frasca is with Univ.\ Grenoble Alpes, CNRS, Inria, Grenoble INP, GIPSA-lab, 38000 Grenoble, France, {\tt paolo.frasca@gipsa-lab.fr}\,.%
	This work was partly supported by CNRS through grants S2IH PEPS ``MOB'' and 80$\vert$PRIME ``DOOM''. The authors would like to thank all the people that have provided comments on early versions of this work, including Doina Bucur, Alessandro Panconesi, Chiara Ravazzi, Tommaso Venturini, and the participants to the 2019 DBAED conference.}
}

\newcommand{\opi}{\mathrm{o}_\mathrm{usr}}		
\newcommand{\prej}{\mathrm{o}_\mathrm{usr}^0}		
\newcommand{\pos}{\mathrm{p}_\mathrm{art}}		

\newcommand{\inter}{\theta}      
\newcommand{\click}{\mathrm{clk}} 

\newcommand{\timepos}{T_{\scriptscriptstyle +}} 
\newcommand{\timeneg}{T_{\scriptscriptstyle -}} 
\newcommand{\recpos}{\mathrm{r}_{\scriptscriptstyle +}} 
\newcommand{\recneg}{\mathrm{r}_{\scriptscriptstyle-}} 
\newcommand{\accpos}{\mathrm{a}_{\scriptscriptstyle+}} 
\newcommand{\accneg}{\mathrm{a}_{\scriptscriptstyle-}} 

\newcommand{\tmax}{t_{\max}} 	
\newcommand{\aveopi}{\overline{\mathrm{o}_\mathrm{usr}}} 
\newcommand{\avepos}{\overline{\mathrm{p}_\mathrm{art}}} 
\newcommand{\ctr}{\mathrm{ctr}} 

\newcommand{\xx}{\mathbf{x}}
\newcommand{\ff}{\mathbf{f}}
\newcommand{\ffRec}{\ff_\mathrm{R}}

\newcommand{\ratiodiff}{\Delta}

\newcommand{\deltao}{\Delta_\mathrm{usr}^{\pm}}		%
\newcommand{\gammactr}{\Gamma_\mathrm{ctr}^{\pm}}		%

\begin{document}
\maketitle 

\begin{abstract} 
In online platforms, recommender systems are responsible for directing users to relevant contents. In order to enhance the users' engagement, recommender systems adapt their output to the reactions of the users, who are in turn affected by the recommended contents. 
%
In this work, we study a tractable analytical model of a user that interacts with an online news aggregator, with the purpose of making explicit the feedback loop between the evolution of the user's opinion and the personalised recommendation of contents. 
More specifically, we assume that the user is endowed with a scalar opinion about a certain issue and seeks news about it on a news aggregator: this opinion is influenced by all received news, which are characterized by a binary position on the issue at hand.  The user is affected by a confirmation bias, that is, a preference for news that confirm her current opinion. The news aggregator recommends items with the goal of maximizing the number of user's clicks (as a measure of her engagement): in order to fulfil its goal, the recommender has to compromise between exploring the user's preferences and exploiting what it has learned so far.
After defining suitable metrics for the effectiveness of the recommender systems (such as the click-through rate) and for its impact on the opinion, we perform both extensive numerical simulations and a mathematical analysis of the model. 
We find that personalised recommendations markedly affect the evolution of opinions and favor the emergence of more extreme ones: the intensity of these effects is inherently related to the effectiveness of the recommender. We also show that by tuning the amount of randomness in the recommendation algorithm, one can seek a balance between the effectiveness of the recommendation system and its impact on the opinions.
\end{abstract}



\section{Introduction}

Recommendation systems are ubiquitous in all kinds of web services, such as search engines, social networking service, e-commerce platforms. 
Their purpose is sieving the information available to them and provide the user with the most relevant items. Recommendation systems leverage a wide array of machine learning techniques, which allow not only to quantify the absolute relevance of the items (like for the PageRank algorithm) but also to tailor the recommendations to the expected tastes of the users, whose online behaviors are suitably recorded. Besides being monumental achievements of computer science, recommendation systems are essential to the user experience and access to contents, information, news or purchase opportunities. 
As online activities become more and more prominent in the lives of the people, questions are asked about the effects (if any) of recommendation systems on the online and offline behaviors of the users. Our investigation specifically questions the role of personalization.

These issues are specially perceived as relevant when it comes to the access to news. Indeed, also news media show a clear trend towards personalised information access. A number of personalised news aggregators like Google News and Yahoo News have emerged: these services collect pieces of news from several media outlets, rank them according to the estimated preferences of the viewing user and propose her the curated collection \cite{Billsus:2000:adaptive,KJJ:2018:survey}. Users can explicitly set their interests and the recommender systems also automatically construct profiles using previous reading patterns~\cite{Liu:2010:PNR,Beam:2014:automating}.  In recommendation systems, personalisation enhances user experience, political activists and scholars have raised concerns that excessive personalisation narrows down the positions available to users about specific issues, effectively enclosing users into so-called ``filter bubbles'' that favour the emergence of opinion polarisation and radicalisation \cite{pariser2011filter,lazer2015rise}. Even though this concern has been downplayed by subsequent research~\cite{Bakshy:2015:exposure}, it is clear that personalization has at least the potential to reinforce the user's idiosyncrasies and biases. Indeed, extensive research has shown that individuals are prone to be affected by a \textit{confirmation bias}. 
By this term we mean the unintentional tendency to acquire and process evidence that confirms one's preconceptions and beliefs, possibly leading to an unconscious one-sided case-building process \cite{nickerson1998confirmation,Mullainathan:2005:market,Quattrociocchi:2017:modeling}. Since empirical evidence supports the idea that confirmation bias is extensive, strong and multiform, its effects may be amplified by curation algorithms.

The main purpose of this paper is to propose a tractable mathematical model of the interplay between a user and a learning system that provides her with personalized recommendations. This model shall allow us to quantify the reciprocal reinforcement of confirmation bias and personalized curation. 
More specifically, we mathematically model the opinion formation process of a user that reads news from a news aggregator that provides personalized recommendations. We restrict ourselves to news that bear implications for one specific issue, say, highlighting the benefits/drawbacks of immigration or arguing in favor/against the impeachment of President Trump. News are characterized by a (binary) attribute that defines their positive or negative {\em position} on the given issue.
The opinion of the user evolves as an affine system that integrates the received news (actually, their positions) along time. Owing to the confirmation bias, news items are clicked upon with a probability that is larger when their position is closer to the user opinion.  
The recommender system has the objective of improving the engagement of the users, measured as the number of clicks. In order to achieve this purpose, the recommender follows a randomized strategy that balances ``exploration'', that is, identifying which position is more appreciated by the user, with ``exploitation'', that is, providing the user with news that are most likely to be clicked on.

In view of our intention to obtain a tractable model, both user and recommender are very stylised. 
However, we believe that our model includes the key features and phenomena that are found in reality: users assimilate the information they receive and are prone to follow their confirmation biases; recommendation systems record the actions of the users and personalize their recommendations to increase the users' engagement. Further discussion on the soundness of our assumptions is provided in Section~\ref{sect:themodel} after describing the model.

We perform extensive simulations of our model and complement them with analytical results.
Our results show that the combination of personalisation and confirmation bias typically makes opinions more extreme. Moreover, more extreme opinions contribute positively to engagement, and therefore to the benefit of the recommender system. As a metter of fact, the performance of the algorithm (in terms of improving the click-through rate) is entangled with its impact on the opinions. The effects of the recommender system can be mitigated by increasing its level of randomness, but this choice also reduces its performance, thereby introducing a relevant trade-off.

\paragraph{Literature review}
Our paper 
is related to several recent works that have addressed recommendation systems or opinion dynamics by mathematical models.
We survey some of the literature that has inspired our work, trying to emphasize similarities and differences. 

A large literature has developed and studied mathematical models of opinion evolution. In this literature, opinions are cognitive orientation towards some objects,  displayed attitudes, or  subjective certainties of beliefs, and as such they can be quantified \cite{friedkin2011social-book}\cite{CFL:2009:stat-phys,AP-RT:17,AP-RT:18}. 
Individuals revise their opinions following social interactions \cite{moussaid2013social} or after obtaining new information, that might confirm or challenge their views. 
Nowadays, much of the social interactions, shopping, information seeking and entertainment activities happen online: experimental studies have demonstrated that online contents can influence feelings and offline behaviours \cite{Bond:2012:experiment,Aral:2012:poked, Kramer:2014:emotional-contagion}.

Motivated by these facts, several recent papers have tried to incorporate some models of online platforms in models of opinion dynamics. The recent paper \cite{PB-PC-GDN:19} models opinion evolution on a social media platform that is able to favour the circulation of certain opinions over others: this effect is obtained by tuning the diffusion probability of the different opinions in the network. Even if the recommender system is not explicitly modelled and there is no personalisation, the study exemplifies the potential distortions introduced by an altered information flow. 
%
In the filter bubble perspective, a potential effect of a platform could be restricting the interactions of the users to a limited number of most similar individuals: our recent paper \cite{RF:2018:cdc-subm} investigates a simple opinion model based on this idea.

Coming to papers with a more defined focus on recommendation systems, \cite{FH:2009:impact-recommender,BLPRT:2016:limit-recommendations} look at systems that recommend products for purchase to a population of users. Both works aim at identifying potential distortions due to the recommender systems, by studying the evolution of the popularities of the products, but bear some relevant differences. Firstly, \cite{BLPRT:2016:limit-recommendations} includes social ties between the users: these ties are shown to mitigate the effects of the recommender. Secondly, in 
 \cite{BLPRT:2016:limit-recommendations} the probability that a product is recommended is proportional to its popularity, whereas in  \cite{FH:2009:impact-recommender} that probability is non-linearly increasing in the popularity. As a consequence, only in the latter model the recommender system is able to distort the market and create hits. This scenario is consistent with what we observe in our model, which is also non-linear in nature.  Similar models of the dynamics of recommenders and socially connected users are studied in the recent paper~\cite{NP-LR:19} by means of simulations.
%

The paper \cite{DGL:2013:goel} investigates the polarising effect on user opinions of collaborative recommender systems used to provide personalised suggestions of items. The authors assume that the items (books in their example) have a binary attribute of which the recommender system is completely agnostic. 
The paper compares three popular recommendation algorithms and 
analytically computes the probability that the next recommended item holds a specific attribute. 
The authors interpret the attribute share of the items owned by the user as her opinion and define as polarising an algorithm that suggests items that reinforce the existing opinion of the user, i.e., that make the attribute share more uneven. In comparison with~\cite{DGL:2013:goel}, which focuses on the effects of a single update step, we rather rigorously study the whole dynamical process including its long-time behavior and the feedback effect.
The paper \cite{Spinelli:2017:cof} adopts the same setting as \cite{DGL:2013:goel}, but investigates numerically the co-evolution of attribute shares and recommendations over a sequence of time steps. 

\paragraph{Outline.}
We conclude this Introduction by an overview of the structure of this paper. In Section~\ref{sect:themodel}, we spell out the detailed description of our mathematical model that includes both the user and the recommender systems. We also discuss the experimental and theoretical backgrounds of our modeling choices. 
Section~\ref{sect:analysis} contains the detailed analysis of the model, which exploits both extensive numerical simulations and mathematical analysis.
In Section~\ref{sect:outro}, we summarize our main results and indicate some directions  for future research. 
An appendix is devoted to exploring the dependance of the results on the model parameters. 


\section{Model: User \& recommender in interaction}\label{sect:themodel}
\begin{figure}
\centering
\begin{tikzpicture} [line width=1.2]

	\node [draw,MatlabGreen] (RS) at (4,1) {%
    	\begin{tabular}{c}{\bf Recommender System} \\[4pt]
	    {\color{black} $\recpos(t)$, $\accpos(t)$, $\recneg(t)$, $\accneg(t)$} \end{tabular}};
	
	\node [draw,MatlabBlue] (user) at (4,3) {
    	\begin{tabular}{c}{\bf User }\\[4pt]
    	{\color{black}opinion $\opi(t)$} \end{tabular}};
	
	\draw [-] (user) -- (7,3) -- (7,2);
	\draw [->,>=latex] (7,2) -- (7,1) -- (RS);
	
	\draw [-] (RS) -- (1,1) -- (1,2);
	\draw [->,>=latex] (1,2) -- (1,3) -- (user);
	
	\node [above left] at (1,1){$\pos(t)$};
	
	\node [below right] at (7,3){$\click(t)$};
	
\end{tikzpicture}
\caption{\label{fig:model-scheme} The closed loop between the user and the news aggregator. The diagram includes the variables exchanged by between the two interacting dynamical systems, as well as their internal state variables. Notation is summarized in Table~\ref{tab:notation}.}
\end{figure}
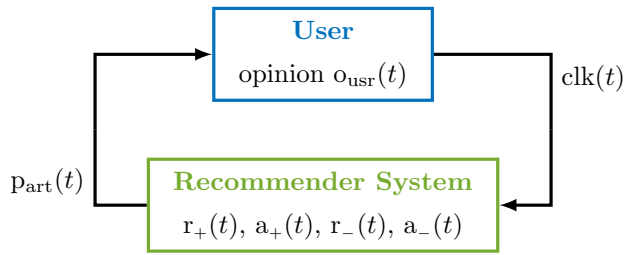

Our purpose is to mathematically model a user that interacts in closed loop with an online news aggregator, see Figure~\ref{fig:model-scheme}. 
The model that we are going to present is therefore made of two components: the {\em user model}, which includes the opinion dynamics and the confirmation bias, and the {\em news aggregator model} with the idealised recommender system. 
The user is endowed with a scalar signed opinion about a specific issue and receives news regarding the issue from an online news aggregator. 
The news aggregator proposes articles to the user, distinguishing between two antithetic positions (positive vs negative). 
%
The news aggregator adopts a recommender system to choose the articles to propose in a personalised way: the system tracks the clicks on the different headlines to understand user's preference and maximise her engagement, i.e., the number of clicks

\subsection{User model: opinion dynamics \& click model}

The user is endowed with a scalar \emph{opinion} that evolves in discrete time  
$$\opi:\N\to [-1,1]\,,$$
and that represents its inclination about a given issue, as well as with a time-independent \emph{prejudice} $\prej \in [-1,1]$ that coincides with her initial opinion about the issue, i.e., $\opi(0)=\prej$.
The prejudice encodes preexisting beliefs, acquired from previous experiences or views of trusted parties. 

The user receives at each time step a news item, that is, an article, that supports a definite 
 \emph{position} about the issue at hand: the position can therefore take on opposite binary values
$$\pos:\N \to \{-1,\plusone\}\,.$$ 
At each time $t$, the user receives an article with position $\pos(t)$ and, 
upon receiving the recommended item, updates her opinion $\opi(t)$ to
\be\label{eq:opinion-model}
\opi(t+1) = \alpha \prej + \beta \opi(t) + \gamma \pos(t) \,,
\ee
where $\alpha,\beta,\gamma$ are non-negative real scalars and $\alpha+\beta+\gamma=1$ (that is, $\opi(t+1)$ is a convex combination of $\prej$, $\opi(t)$ and $\pos(t)$).
%
The weights $\alpha, \beta$ and $\gamma$ describe the relative importance of the prejudice, of the previous opinion (memory) and of the new information, respectively, in shaping the user's new opinion.


%
%
%
%
%
%
%
%
%
%
%
%
%

\begin{figure}[!t]
\centering
\boxfig{\includegraphics[page =1, trim={\COMPfigtriml} {\COMPfigtrimb} {\COMPfigtrimr} {\COMPfigtrimt}, clip ,width={\COMPfigwidth}, keepaspectratio=true]{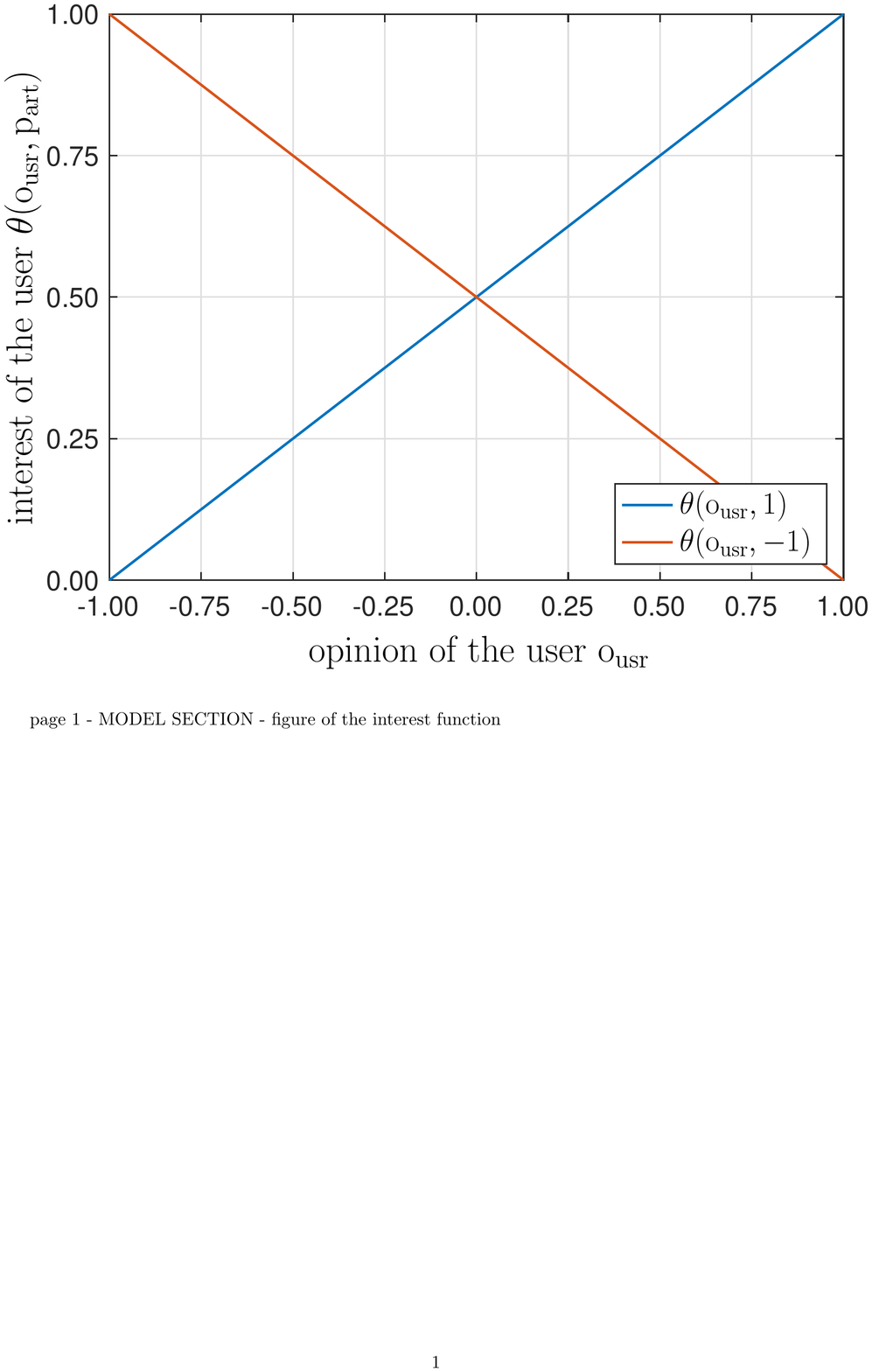}}
\caption{\label{fig:click_function}The functions $\inter(\opi,\pos)$ as in~\eqref{eq:g-fun-example} with $\pos=\plusone$ and $\pos=-1$.}
\end{figure}

Beside updating her opinion, the user decides whether to read the recommended article or not, i.e., to click on its headline or not, following her interest. 
We model the $\{0,1\}$ \emph{click decision}, meaning $\{\text{no click}, \text{click}\}$, as a stochastic process such that at each time the decision is a Bernoulli random variable with an opinion-dependent parameter $\inter(\opi,\pos)$, i.e., 
$$\click(t) \sim {\rm Bernoulli}(\inter(\opi(t),\pos(t)))\,.$$ 
The function 
$\inter: [-1,1]\times\{-1,\plusone\} \to [0,1]$  quantifies the subjective interest $\inter(\opi,\pos)$ of the user with opinion $\opi$ into an item with headline of position $\pos$. 
%
The user is subject to a confirmation bias \cite{nickerson1998confirmation} and prefers contents 
consistent with her opinion $\opi$. To model this fact, we take the function 
\be\label{eq:g-fun-example}
\inter(\opi,\pos) = \frac12 + \frac12 \,\opi\,\pos \,,
\ee
which is depicted in Figure~\ref{fig:click_function}.

\paragraph{On the justification of the user model.}
In our model, opinions belong to the interval $[-1,1]$, which are sometimes referred to as {\it polar} opinions and indicate the degree of proclivity towards one or two competing alternatives~\cite{VA-FB-KS:17}. The opinion dynamics assimilate the information that the user receives. The specific linear model~\eqref{eq:opinion-model} is very close to various models proposed both by sociologists and economists in the context of naive learning~\cite{BG-MOJ:10,DA-AO:11} and of opinion formation under social influence~\cite{friedkin1990social}.
Assuming that the evolution follows a convex (and therefore, positive) combination is consistent with observations in experimental social psychology, for both live \cite{Friedkin:2015:css} and online interactions~\cite{BCH-MM-BP:17}, that post-discussion opinions get closer than pre-discussion opinions. In our case, the stream of news items is the source of information and is assimilated by the user~\cite{BG-MOJ:10}.  

In modeling the effect of the news item, we make a twofold assumption: we assume that the article's position about the issue is binary in nature and that 
the user revises her opinion upon being recommended the item, that is, without the need to read the full article. In fact, we assume that the article's headline is sufficient to influence the user to revise her opinion. 
The binary nature of the items corresponds well to situations where news outlets take partisan perspectives: it has been observed that newspapers can slant the presentation of news to match their audience' preferences \cite{Mullainathan:2005:market}.
The assumption that receiving the headline is sufficient to be influenced is consistent with the heuristic model of persuasion by \cite{chaiken:1987:heuristic}, based on the observation that opinion changes are often the outcome of minimal amounts of information and superficial judgements.
There is however another, perhaps more compelling, reason to make this assumption: this assumption allows us to focus more precisely our analysis on the effect of closed-loop recommendations, in the following sense. Since the user is influenced by all received news items, she effectively has no confirmation bias as per her process of information assimilation and opinion update, whereas her bias is made apparent to the recommender system through her clicking history. Informally speaking, we may say that in our model any bias in the dynamics of the user opinion shall be due to the effect of {\em biased recommendations}. Indeed, unbiased recommendations would result in an unbiased opinion and the confirmation bias only bears effects through the recommendation system. Our assumption is therefore conservative in nature: assuming the user to be influenced only after reading the articles can only result in a stronger bias on the opinion.

Finally, we note that the specific expression \eqref{eq:g-fun-example} comes from the literature as it translates into our framework the definition of {\em biased user} by \cite[Def. 9]{DGL:2013:goel}. Moreover, we observe that it satisfies several common-sense properties:
\begin{itemize}
\item $\inter(\opi,\plusone) \ge \inter(\opi,-1)$ for $\opi>0$ while $\inter(\opi,\plusone) \le \inter(\opi,-1)$ for $\opi<0$: the interest of the user is higher for articles that have a position closer to her opinion;
\item $\inter(\opi,\plusone)$ is non-decresing in $\opi$ while $\inter(\opi,-1)$ is non-increasing in $\opi$: if the match between the opinion of the user and the position of the article increases, the interest of the user does not decrease; 
\item $\inter(\opi,\plusone) = \inter(-\opi,-1)$: the interest is symmetric in the opinion-position match;
\item $\inter(\plusone,\plusone) = \inter(-1,-1) = 1$: complete alignment between opinion and position makes the click almost certain. 
\end{itemize}
%

%

\subsection{News aggregator model: $2\eps$-greedy recommender system}

The news aggregator has the purpose of \emph{maximising} the \emph{user engagement}, measured by the \emph{ratio of clicks} on the suggested contents. At each time step the news aggregator has to choose whether to recommend an article with ``positive'' or ``negative'' position. 
The online service tracks the user's activities by logging clicks, interpreted as positive votes for the positions of the corresponding articles. The recommender system models the user's interest in a position as the probability that the user clicks on an article bearing that position and the predicted interests can be used to generate the recommendation \cite{Liu:2010:PNR}. 
The recommender system faces the \textit{exploration-exploitation dilemma} of sequential decision problems, which arises between staying with the most successful option so far and exploring the other option, which might turn better in the future \cite{bubeck:2012:libro-bandit,LCLS:2010:contextual-bandit-news-recommendation}. 
Moreover, as users' interests change over time, the system needs to incrementally update the user's profile to reflect such changes.
%
We adopt for the recommender system an {\em $2\eps$-greedy algorithm}: with probability $2\eps$, the recommender system randomly explores the binary options; with probability $1-2\eps$, the recommender system recommends the most successful option so far.
This approach, although not optimal, is supported by the literature for time-varying settings like ours \cite{koulouriotis2008reinforcement}. 

Let us now describe in details how the system keeps track of past clicks to learn the most succesfull choice. The sets $\timepos(t)$ and $\timeneg(t)$ collect the time steps preceding $t$ at which an article with position $\plusone$ or $-1$, respectively, has been recommended to the user
\begin{align*}
&\timepos(t) = \{s: 0\leq s\leq t-1 \text{ and } \pos(s) = \plusone\}\,, \\
&\timeneg(t) = \{s: 0\leq s\leq t-1 \text{ and } \pos(s) = -1\}\,.
\end{align*}
The cardinalities of the above sets
$$\recpos(t) = \# \timepos(t)\,,\qquad \qquad  \recneg(t) = \# \timeneg(t)\,, $$ 
count how many times an article with position $\plusone$ or $-1$ has been proposed until time $t$. 
The counters $\accpos(t)$ and $\accneg(t)$ record how many times until time $t$ the user has accepted a recommendation with positions $\plusone$ and $-1$, respectively, 
$$\accpos(t) = \sum_{s \in \timepos(t)} \click(s)\,,  \qquad \qquad \accneg(t) = \sum_{s \in \timeneg(t)} \click(s)\,.$$
According to the previous discussion, we shall indeed evaluate the performance of the recommender systems by the {\em click-through rate} 
\begin{equation}\label{eq:ctr-def}\ctr(t) = \frac{\accpos(t) +  \accneg(t)}{t}.\end{equation}

We observe that,  by their definition, the integer sequences $\recpos(t)$, $\recneg(t)$, $\accpos(t)$ and $\accneg(t)$ have null initial value 
and 
evolve according to 
$$\arraycolsep=2pt 
\left\{ \begin{array}{ll}
    \text{if ~}\pos(t)= \plusone \text{~ then ~}& 
    \left\{\begin{array}{lclclcl} 
        \recpos(t+1) &=& \recpos(t)+1	\,,     &\quad& \accpos(t+1) &=& \accpos(t) + \click(t)\,, \\
        \recneg(t+1) &=& \recneg(t)		\,, 	&\quad& \accneg(t+1) &=& \accneg(t)\,;
    \end{array}\right. \\[15pt]
    \text{if ~}\pos(t) = -1 \text{~ then ~}& 
    \left\{\begin{array}{lclclcl} 
        \recpos(t+1) &=& \recpos(t)		\,,     &\quad& \accpos(t+1) &=& \accpos(t)\,,\\
        \recneg(t+1) &=& \recneg(t)+1	\,, 	&\quad& \accneg(t+1) &=& \accneg(t)+\click(t)\,.
    \end{array}\right.
\end{array} \right.$$

%
At any time $t\ge 2$ the recommender system computes the ratios $\frac{\accpos(t)}{\recpos(t)}$ and  $\frac{\accneg(t)}{\recneg(t)}$, i.e., the proportions of clicks collected by the positions $\plusone$ and $-1$, respectively, to   estimate the future odds of collecting a click upon proposing the positions $\plusone$ and $-1$. Then the recommender system proposes with probability $1-\eps$ an article representing the position that has received the higher proportion of clicks and with the complementary probability $\eps$ the other position. Should a tie arise, both positions receive equal probabilities. 
In formulas, the decision rule reads 
\be \label{eq:recommender-rule}
\left\{\begin{array}{lll}
\text{if ~} \frac{\accpos(t)}{\recpos(t)} > \frac{\accneg(t)}{\recneg(t)} \text{~ then}\quad & \P(\pos(t) = \plusone) =1 - \eps, \quad & \P(\pos(t) = -1) =\eps  \\[4pt]
\text{if ~} \frac{\accpos(t)}{\recpos(t)} = \frac{\accneg(t)}{\recneg(t)} \text{~ then}\quad & \P(\pos(t) = \plusone) =0.5,\quad& \P(\pos(t) = -1) =0.5  \\[4pt]
\text{if ~} \frac{\accpos(t)}{\recpos(t)} < \frac{\accneg(t)}{\recneg(t)} \text{~ then}\quad  & \P(\pos(t) = \plusone) =\eps,\quad &\P(\pos(t) = -1) =1-\eps \,.
\end{array}\right.
\ee
The parameter $\eps$ controls the trade-off between exploration and exploitation and is typically small; in most numerical simulations we shall take $\eps = 0.05$. At times $0$ and $\plusone$ the recommender system follows an initialisation procedure  where both positions are proposed in a random order, i.e., 
$$\P(\pos(0) = \plusone,\, \pos(1)=-1) = 0.5\,,\qquad\P(\pos(0) = -1,\, \pos(1)=\plusone) = 0.5 \,, $$
to make sure that $\recpos(2) = \recneg(2) = 1$.

%

\section{Analysis of the closed-loop dynamics}\label{sect:analysis}
We now move on to present our findings about the dynamical model described in the previous section. Our presentation is divided into two main parts. In the first part, we observe that typical trajectories of the dynamical model are characterized by a definite majority of either $+1$ or $-1$ recommendations. This observation supports the study of the expected dynamics {\em conditioned upon a given majority}: these conditional expectations can be derived in closed-form and turn out to describe the stochastic dynamics very accurately.
In the second part, we build on these formal derivations and on extended simulations to discover the effects of recommendations on the evolution of the opinions.
Before these developments, we formally write down the closed-loop dynamics that summarizes the model in the previous section. 
\renewcommand{\arraystretch}{1.0}
\begin{table}
\centering
\begin{tabular}{  l  p{6.0cm} l}
    \hline			
    \textbf{Symbol} & \textbf{Meaning} & \textbf{Notes} \\
    \hline
    
    $\opi(t)$		 				& user opinion 							&\\
    $\prej$ 						& user prejudice						& \\
    $\pos(t)$ 						& article position 						&  \\
    $\alpha,\beta,\gamma$ 			& opinion model parameters 				& See Equation \eqref{eq:opinion-model} \\
    $\inter(\opi,\pos)$ 			& interest parameter 								&  \\
    $\click(t)$ 					& click (Bernoulli random variable)					&  \\
    $\recpos(t)$, $\recneg(t)$ 		& number of articles with position $\plusone$ or $-1$, respectively, recommended before time $t$		&  \\
    $\accpos(t)$, $\accneg(t)$ 		& number of articles with position $\plusone$ or $-1$, respectively, accepted before time $t$		&   \\
    $\epsilon$ 						& recommender system parameter 						&  \\
    
    %
    $\tmax$ 			& simulation final time 				&  \\
    $\aveopi(t) $  		& time averaged user opinion   			&  $\aveopi(t) = \frac1{t} \sum_{s=0}^{t-1} \opi(s)$  \\
    $\avepos(t) $ 		& time averaged article position 	 	& $\avepos(t) = \frac1{t} \sum_{s=0}^{t-1} \pos(s)$ \\
    $\ctr(t) $			& click-through rate					& $\ctr(t) = \frac{\accpos(t)+ \accneg(t)}{t} $ \\

    $\xx$				& system state							&  $\xx = [\recpos, \recneg, \accpos, \accneg, \opi]^\top$	 \\
    $A$  				& system matrix (linear part)			&  See Equation \eqref{eq:system-compact-form} \\
    $\ff(\xx)$ 			& system additive non-linear part 		&  \\
    $\ratiodiff(\xx)$ 	& ratio difference 						& Equation~\eqref{eq:ratiodiff-def}\\
    $h_\epsilon(z)$ 	& modified step function of scalar $z$	&  		\\

	$\deltao$			& opinion distortion					& See Equation \eqref{eq:deltao} \\
	$\gammactr$			& click-through rate gain				& See Equation \eqref{eq:deltactr} \\
    \hline   
\end{tabular}
  \caption{Summary of the notation used in the paper. 
 } \label{tab:notation}
\end{table}

By defining  the state vector 
$\xx(t):= [\recpos(t), \recneg(t), \accpos(t), \accneg(t), \opi(t)]^\top\,$,
the dynamics\footnote{For times $t=0,1$ the system is driven by the stochastic initialization of the recommender system. The specific form of $\xx(2)$ can be derived but is not relevant to our analysis that focuses on long-term behaviors. We therefore avoid to report it.
}
  can indeed be written as
\begin{equation} \label{eq:system-compact-form}
\left\{ \begin{array}{l} 
\xx(t+1) = A \xx(t) + \ff(\xx(t)) \qquad t\ge2 \\
\xx(2) \text{ discrete random variable\,,}
\end{array}\right. \end{equation}
where 
$A$ is the update matrix 
\begin{equation*}
A = \begin{bmatrix}
    1 & 0 & 0 & 0 & 0  \\
    0 & 1 & 0 & 0 & 0  \\
    0 & 0 & 1 & 0 & 0  \\
    0 & 0 & 0 & 1 & 0  \\
    0 & 0 & 0 & 0 & \beta  \\
  \end{bmatrix}\,
\end{equation*}
and 
the vector $\ff(\xx(t))$ 
is a random variable that can take on four values, corresponding to the cases
 ``position $\plusone$, no click'', 
  ``position $\plusone$, get click'', 
   ``position $-1$,  no click'' and
    ``position $-1$,  get click''.
In the recommender rule \eqref{eq:recommender-rule}, the probability of each of these cases depends non-linearly on the state $\xx(t)$. 
To encode these probabilities, we define the difference 
\be \label{eq:ratiodiff-def}
\ratiodiff(\xx(t)) := 
\frac{\accpos(t)}{\recpos(t)} - \frac{\accneg(t)}{\recneg(t)}\,.
\ee
Clearly, $\Delta(x(t))>0$ if and only if +1 is the most succesfull recommendation so far. Using the modified step function $h_\eps:\R\to \{\eps, \frac12, 1-\eps\}$  
$$h_\eps(x) := \left\{ \begin{array}{ll} 
	1-\eps & \text{if } x>0 \\[2pt] 
	\frac{1}{2} & \text{if } x = 0\\[2pt] 
	\eps & \text{if } x<0 \,,
\end{array}\right. $$
the 
 probabilistic model of $\ff(\xx)$ is 
$$ \ff(\xx) = \left\{\begin{array}{lll}
[1, 0, 1, 0, \alpha \prej + \gamma]^\top  &\text{ with probability }  & h_\eps(\ratiodiff(\xx)) \,  			\inter(\opi,\plusone) \\[4pt]
[1, 0, 0, 0, \alpha \prej + \gamma]^\top  &\text{ with probability }  & h_\eps(\ratiodiff(\xx)) \,  			\big(1-\inter(\opi,\plusone) \big) \\[4pt]
[0, 1, 0, 1, \alpha \prej - \gamma]^\top  &\text{ with probability }  & \big(1-h_\eps(\ratiodiff(\xx))\big) \,	\inter(\opi,-1) \\[4pt]
[0, 1, 0, 0, \alpha \prej - \gamma]^\top  &\text{ with probability }  & \big(1-h_\eps(\ratiodiff(\xx))\big) \,	\big( 1-\inter(\opi,-1) \big).
\end{array}\right.$$



\subsection{Majority trajectories: numerical evidences and  analysis}

\begin{figure}
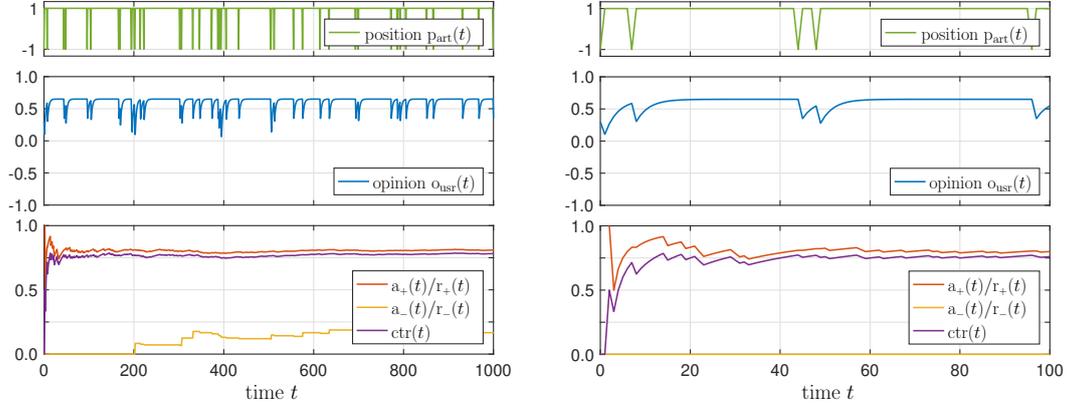

\centering
\boxfig{\includegraphics[page =4, trim={\COMPfigtriml} {\COMPfigtrimbX} {\COMPfigtrimr} {\COMPfigtrimt}, clip ,width={\COMPfigwidth}, keepaspectratio=true]{compiled_figures.pdf}}
\boxfig{\includegraphics[page =5, trim={\COMPfigtriml} {\COMPfigtrimbX} {\COMPfigtrimr} {\COMPfigtrimt}, clip ,width={\COMPfigwidth}, keepaspectratio=true]{compiled_figures.pdf}}
\caption{\label{fig:Dyn_basic_up}Simulation with $\alpha = 0.15$, $\beta = 0.70$, $\gamma = 0.15$, $\prej = 0.30$ and recommended choice of $\pos(t)$ with $\epsilon = 0.05$. 
Left: up to time $\tmax = 1000$. Right: the first 100 steps. 
The most recommended position is $\plusone$.}
\end{figure}

\begin{figure}
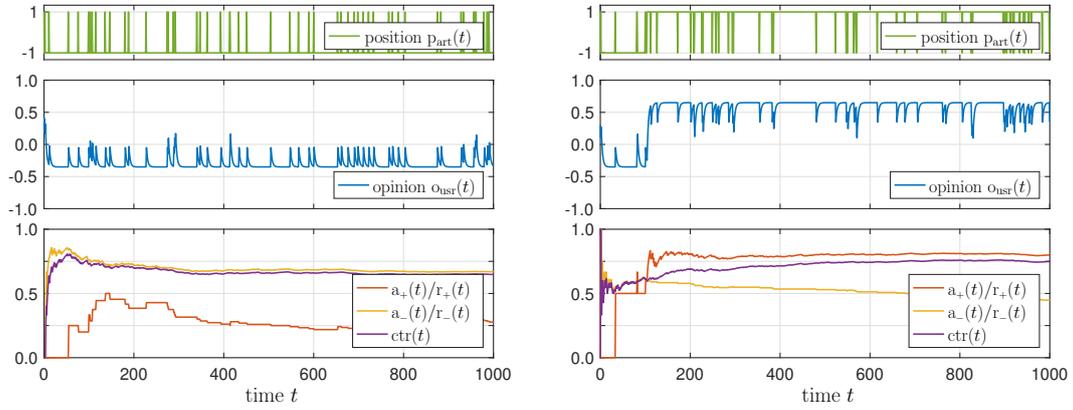

\centering
\boxfig{\includegraphics[page =6, trim={\COMPfigtriml} {\COMPfigtrimbX} {\COMPfigtrimr} {\COMPfigtrimt}, clip ,width={\COMPfigwidth}, keepaspectratio=true]{compiled_figures.pdf}}
\boxfig{\includegraphics[page =8, trim={\COMPfigtriml} {\COMPfigtrimbX} {\COMPfigtrimr} {\COMPfigtrimt}, clip ,width={\COMPfigwidth}, keepaspectratio=true]{compiled_figures.pdf}}
\caption{\label{fig:due-esempi}Simulations with $\alpha = 0.15$, $\beta = 0.70$, $\gamma = 0.15$, $\prej = 0.30$ and recommended choice of $\pos(t)$ with $\epsilon = 0.05$. 
Left: a $-1$-majority trajectory. Right: a $\plusone$-majority trajectory that features a majority of $-1$s in its initial phases.}
\end{figure}

\begin{figure}
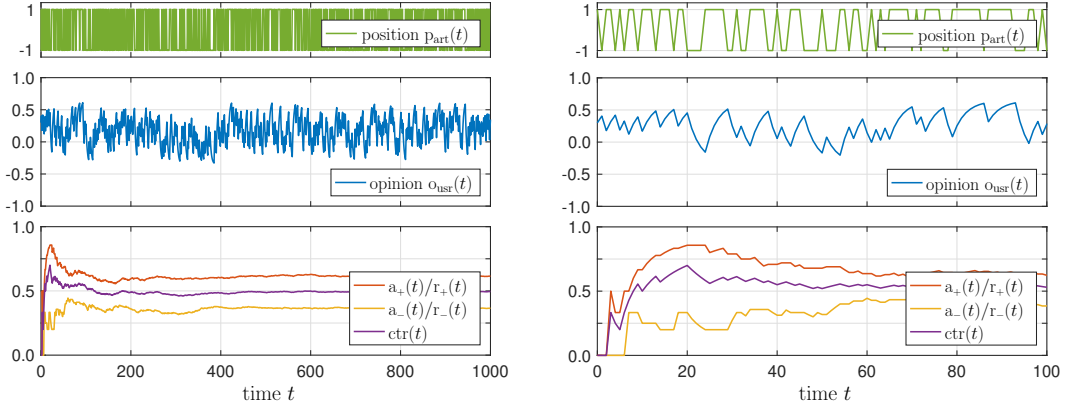

\centering
\boxfig{\includegraphics[page =2, trim={\COMPfigtriml} {\COMPfigtrimbX} {\COMPfigtrimr} {\COMPfigtrimt}, clip ,width={\COMPfigwidth}, keepaspectratio=true]{compiled_figures.pdf}}
\boxfig{\includegraphics[page =3, trim={\COMPfigtriml} {\COMPfigtrimbX} {\COMPfigtrimr} {\COMPfigtrimt}, clip ,width={\COMPfigwidth}, keepaspectratio=true]{compiled_figures.pdf}}

\caption{\label{fig:Dyn_basic_no_algo}Simulation with $\alpha = 0.15$, $\beta = 0.70$, $\gamma = 0.15$, $\prej = 0.30$ and random choice of $\pos(t)$ (i.e $\epsilon = 0.50$). 
Left: up to time $\tmax = 1000$. Right: the first 100 steps.}
\end{figure}

To showcase the evolution of the closed-loop dynamics~\eqref{eq:system-compact-form}, we present in detail a set of simulations in Figures~\ref{fig:Dyn_basic_up}, \ref{fig:due-esempi} and \ref{fig:Dyn_basic_no_algo}. 
%
%
An immediate observation is that the dynamics are shaped by a prevalence of $+1$ or $-1$ recommendations. For instance, the simulation in Figure~\ref{fig:Dyn_basic_up} is clearly characterized by having $\pos(t)=\plusone$ most of the times. Correspondingly, the ratio ${\accpos(t)}/{\recpos(t)}$ is larger and more stable than the ratio ${\accneg(t)}/{\recneg(t)}$. These facts are consistent with the opinion $\opi(t)$ being always positive (often around $0.65$) and $\plusone$ being by far the most recommended position. We refer to this feature as being a {\em \up\ trajectory}. Conversely, Figure~\ref{fig:due-esempi} (left) shows a \down\ trajectory. 
%
%
To highlight the effects of the recommender system, 
these dynamics should be compared against Figure~\ref{fig:Dyn_basic_no_algo}, where 
the sequence of recommended positions is purely random, that is, $\eps = \frac12$. We can observe that in this case the click-through rate $\ctr(t)$ is much smaller than in Figure~\ref{fig:Dyn_basic_up}.  


With the support of these initial insights from the simulations, we proceed to develop a mathematical analysis of the stochastic dynamics \eqref{eq:system-compact-form}.
To this purpose, it is natural to calculate the conditional expectation 
\be\label{eq:expect-difficult}
\E[\xx(t+1)|\xx(t)]  = \E[A\xx(t)+ \ffRec(\xx(t))|\xx(t) ] 
    = \begin{bmatrix}
    \recpos(t)    \\ 
    \recneg(t)  \\ 
    \accpos(t) \\ 
    \accneg(t)  \\ 
    \beta \opi(t)  
\end{bmatrix}    + \begin{bmatrix}
      h_\eps(\ratiodiff(\xx(t)))  \\ 
      1-h_\eps(\ratiodiff(\xx(t))) \\ 
      h_\eps(\ratiodiff(\xx(t))) \,\inter(\opi(t),\plusone) \\ 
      \big(1-h_\eps(\ratiodiff(\xx(t)))\big)\,\inter(\opi(t),-1) \\ 
      \alpha \prej + \gamma \big(2h_\eps(\ratiodiff(\xx(t))) -1\big)
\end{bmatrix}
\ee
%
for $t\ge 2$. 
In principle, one could now derive the evolution of  $\E[\xx(t)]$ by taking the expectation of both sides: however, we refrain from this operation for two reasons.  Not only the calculation is made difficult by the function $h_\eps(\ratiodiff(\xx))$ being non-linear, but is also unlikely to bring significant insight. Indeed, trajectories are distinctly dominated either by either $+1$ or $-1$ recommendations: therefore the average dynamics would not describe well these two distinct groups. Motivated by this observation, we instead study the process conditional on having either $\ratiodiff(\xx(t))>0$ or $\ratiodiff(\xx(t)) <0$ for $t\ge2$. 
%
%
We thus introduce the notation
\begin{equation}\label{eq:condizionamenti}\Ep[\xx(t)]:=\E[\xx(t)\,|\,\ratiodiff(\xx(s))>0 \text{~ for } 2\le s<t]\,,\quad 
	\Em[\xx(t)]:=\E[\xx(t)\,|\,\ratiodiff(\xx(s))<0 \text{~ for } 2\le s<t] \end{equation}
and we proceed with the calculations for case  $\ratiodiff(\xx(t))>0$, which means $h_\eps(\ratiodiff(\xx(t))) = 1-\eps$. We recall that in this scenario the most favourable option is recommending articles with position $\plusone$. Equations \eqref{eq:expect-difficult} and \eqref{eq:g-fun-example} give 
\begin{equation*}
\Ep[\xx(t+1)|\xx(t)] = 
    \begin{bmatrix}
    \recpos(t)     \\ 
    \recneg(t)   \\ 
    \accpos(t)   \\ 
    \accneg(t)   \\ 
    \beta \opi(t)  
    \end{bmatrix} + \begin{bmatrix}
     1-\eps  \\ 
      \eps \\ 
     (1-\eps) \,\inter(\opi(t),\plusone) \\ 
     \eps\,\inter(\opi(t),-1) \\ 
     \alpha \prej + \gamma \,(1-2\eps)  
\end{bmatrix}
=\begin{bmatrix}
  \recpos(t)  + 1-\eps \\ 
  \recneg(t)  + \eps  \\ 
  \accpos(t)   + \frac12(1-\eps)   \opi(t) + \frac12(1-\eps) \\ 
  \accneg(t)   -\frac12 \eps   \opi(t) + \frac12 \eps  \\ 
\beta\,  \opi(t)  +  \alpha\, \prej + \gamma\,(1-2\eps) 
\end{bmatrix}
\end{equation*}
We take the expectation a second time and get
\be
\Ep[\xx(t+1)]  = \BRp \Ep[\xx(t)] + \bbRp 
 \ee
where 
\begin{equation*}
\BRp = \begin{bmatrix}
    1 & 0 & 0 & 0 & 0  \\
    0 & 1 & 0 & 0 & 0  \\
    0 & 0 & 1 & 0 & \frac12(1-\eps)  \\
    0 & 0 & 0 & 1 & -\frac12 \eps  \\
    0 & 0 & 0 & 0 & \beta  
  \end{bmatrix} \quad\text{ and }\quad 
\bbRp= \begin{bmatrix}
    1-\eps  \\
    \eps  \\
    \frac12(1-\eps)  \\
    \frac12 \eps \\
    \alpha \prej  +\gamma (1-2\eps)
  \end{bmatrix}\,.
\end{equation*}
The quantity $\Ep[\opi(t)]$ has an autonomous dynamics and converges to
\be\label{eq:Ep-z-inf}
 \lim_{t \to \infty} \Ep[ \opi(t)] = \frac{\alpha \prej + \gamma(1-2\eps)}{\alpha+\gamma}.
\ee
%
The calculations for the case $\ratiodiff(\xx(t))<0$, meaning $h_\eps(\ratiodiff(\xx(t))) = \eps$, are analogous and yield
\be\label{eq:Em-z-inf}
 \lim_{t \to \infty} \Em[\opi(t)] = \frac{\alpha \prej - \gamma(1-2\eps)}{\alpha+\gamma}
 \ee
Therefore, the limit values of $\Ep[ \opi(t)]$ and $\Em[\opi(t)]$ differ by  
\be\label{eq:discrepancy}
 \lim_{t \to \infty} \Ep[ \opi(t)] - \lim_{t \to \infty} \Em[\opi(t)] = 
 2\frac{\gamma}{\alpha+\gamma}(1-2\eps), 
\ee
which does not depend on $\prej$. We call \textit{discrepancy} the quantity \eqref{eq:discrepancy}, which we take as a {\em measure of the effect of recommendations on opinions}, and which corresponds to the vertical gap between the green and magenta lines in Figure~\ref{fig:trajectories-split} (right). This figure confirms that {\it trajectories split into $+1$-trajectories and $-1$-trajectories, concentrating around the conditional expectations}.

\paragraph{Recommender dynamics and click-through rate.}
Having understood the evolution of the user dynamics, we can proceed to study the internal variables of the recommender, which follow
\begin{align*}
\Ep[ \recpos(t+1)] -\Ep[ \recpos(t)] &= 1-\eps \\ 
\Ep[ \recneg(t+1)] -\Ep[ \recneg(t)] &= \eps \\ 
\Ep[ \accpos(t+1)] -\Ep[ \accpos(t)] &=  \frac12(1-\eps) + \frac12(1-\eps) \Ep[ \opi(t)]  \\ 
\Ep[ \accneg(t+1)] -\Ep[ \accneg(t)] &=  \frac12 \eps - \frac12 \eps \Ep[ \opi(t)]\,.
\end{align*}
Therefore, for large $t$ we have 
\begin{align}
\lim_{t\to\infty} \frac{\Ep[ \recpos(t)]}t &= 1-\eps  \label{eq:trendpp} \\ 
\lim_{t\to\infty} \frac{\Ep[ \recneg(t)]}t &= \eps  \label{eq:trendpm}\\ 
\lim_{t\to\infty} \frac{\Ep[ \accpos(t)]}t &= \frac12(1-\eps)( 1+\Ep[ \opi(\infty)] ) \label{eq:trendrp}\\ 
\lim_{t\to\infty} \frac{\Ep[ \accneg(t)]}t &= \frac12\eps ( 1-\Ep[ \opi(\infty)] ) \label{eq:trendrm}
\end{align}
and consequently, the expected click-through rate \eqref{eq:ctr-def} becomes
\be \label{eq:Ep-click}
\lim_{t\to\infty} \Ep[\ctr(t)] 
=  \frac12 + \frac12(1-2\eps)\Ep[ \opi(\infty)] 
	= \frac12 + \frac12(1-2\eps)\frac{\alpha \prej + \gamma(1-2\eps)}{\alpha+\gamma}\,.	
	\ee
An analogous calculation gives
\be \label{eq:Em-click}
\lim_{t\to\infty}\Em[ \ctr(t)] =  \frac12 - \frac12(1-2\eps)\Em[  \opi(\infty)]   
	= \frac12 - \frac12(1-2\eps)\frac{\alpha \prej - \gamma(1-2\eps)}{\alpha+\gamma}\,.
\ee

Observe that the click-through rate achieved by the recommender system is larger than the click-through rate with the random choice $\eps=0.5$. 
%
The difference between the asymptotic click-through rates 
is
$$\Ep[ \ctr(\infty)] - \Em[ \ctr(\infty)] =  \frac{\alpha}{\alpha+\gamma}\prej(1-2\eps)\,.$$
This quantity is proportional to $1-2\eps$ and to $\prej$: intuitively, recommending $\plusone$ is more rewarding for larger $\prej$. This last comment leads us to look more carefully at the link between predjudices and recommendations.

\begin{figure}
\includegraphics[page =12, trim={\COMPfigtrimlASY} {\COMPfigtrimbASY} {\COMPfigtrimrASY} {\COMPfigtrimtASY}, clip ,width={\COMPfigwidthASY}, keepaspectratio=true]{compiled_figures.pdf}
\includegraphics[page =13, trim={\COMPfigtrimlASY} {\COMPfigtrimbASY} {\COMPfigtrimrASY} {\COMPfigtrimtASY}, clip ,width={\COMPfigwidthASY}, keepaspectratio=true]{compiled_figures.pdf}

\caption{
The average opinion $\aveopi(\tmax)$ with $\tmax =1000$ corresponding to simulations with $\alpha = 0.15$, $\beta = 0.70$, $\gamma = 0.15$ and $\prej \in \{-1.00, -0.90, \ldots, 1.00\}$.  There are 1000 simulations for each prejudice $\prej$.
Left: random choice of $\pos(t)$ (i.e. $\eps = 0.50$). 
Right: recommended choice of $\pos(t)$ with $\epsilon = 0.05$; \up~and \down~simulations are distinguished; the equations of the magenta and green line are \eqref{eq:Ep-z-inf} and \eqref{eq:Em-z-inf}, respectively.
%
}
\label{fig:trajectories-split}
\end{figure}
\medskip


\paragraph{Strong prejudices lead to according recommendations.}
\begin{figure}
\includegraphics[page =28, trim={\COMPfigtrimlASY} {\COMPfigtrimbASY} {\COMPfigtrimrASY} {\COMPfigtrimtASY}, clip ,width={\COMPfigwidthASY}, keepaspectratio=true]{compiled_figures.pdf}
\includegraphics[page =29, trim={\COMPfigtrimlASY} {\COMPfigtrimbASY} {\COMPfigtrimrASY} {\COMPfigtrimtASY}, clip ,width={\COMPfigwidthASY}, keepaspectratio=true]{compiled_figures.pdf}	

\caption{Parameters: $\alpha = 0.20$, $\beta = 0.70$, $\gamma = 0.10$, $\epsilon = 0.05$. For each prejudice $\prej$ there are 1000 simulations with $\tmax =5000$. Left: the time averaged opinion $\aveopi(\tmax)$, with the simulations distinguished by their \up~and \down~character. The equations of the magenta and green line are \eqref{eq:Ep-z-inf} and \eqref{eq:Em-z-inf}, respectively. 
Right: the empirical probability of \up~and \down~trajectory computed using $\avepos(\tmax)$.
Dashed blue lines have abscissas \eqref{eq:z0-larger} \eqref{eq:z0-smaller}.}\label{fig:multi_selfconsist_example}
\end{figure}
%


Having obtained a clear analytical picture under the assumption that trajectories have a definite majority, we now consider more carefully whether an approach based on \eqref{eq:condizionamenti} is justified. 
%
In this respect, it is comforting that for some combinations of $\alpha, \beta,\gamma, \epsilon$ and $\prej$ one of the two kind of trajectories is extremely unlikely. 
%
%
%
Indeed, in the scenario $\ratiodiff(\xx(t))>0$ we have 
\begin{align*}
\lim_{t\to\infty}\frac{\Ep[ \accpos(t)]}{\Ep[ \recpos(t)]} - \frac{\Ep[ \accneg(t)]}{\Ep[ \recneg(t)]}
=
\frac12(1+  \Ep[ \opi(\infty)]  ) -\frac12( 1-\Ep[ \opi(\infty)] ) 
= \Ep[ \opi(\infty)].  
\end{align*}
Therefore, we should have $\Ep[ \opi(\infty)] >0$, which means that the opinion $\opi(t)$ should be positive. 
This makes sense: due to the expression \eqref{eq:g-fun-example} of $\inter(\opi(t),\pos(t))$, if opinion $\opi(t)$ remains positive, then position $\plusone$ has better odds of collecting clicks than position $-1$ and hence $\ratiodiff(\xx(t))$ gets positive.
Using \eqref{eq:Ep-z-inf} the condition $\Ep[ \opi(\infty)] >0$ reads 
\be\label{eq:z0-larger}
\prej > - \frac\gamma\alpha (1-2\eps)\,,
\ee
i.e., $\prej$  should not be too negative. Reversing the argument, any $\prej$ that satisfies this condition is compatible with $\ratiodiff(\xx(t))>0$ and \up~trajectories. 
In the scenario $\ratiodiff(\xx(t))<0$, analogous calculations lead to
\be\label{eq:z0-smaller}
\prej < \frac\gamma\alpha (1-2\eps)\,.
\ee
Together, conditions \eqref{eq:z0-larger} \eqref{eq:z0-smaller} split the interval $[-1,1]$ of possible prejudices $\prej$ in three parts: 
\begin{enumerate}[label=\Alph*]
\item \label{item:A} $\prej<-\frac\gamma\alpha (1-2\eps)$ : only $\ratiodiff(\xx(t))<0$ seems possible and we should observe only \down~trajectories;
\item \label{item:B} $-\frac\gamma\alpha (1-2\eps)\le \prej \le \frac\gamma\alpha (1-2\eps)$ : both $\ratiodiff(\xx(t))<0$ and $\ratiodiff(\xx(t))>0$ are possible, allowing  both \up~and \down~trajectories;
\item \label{item:C} $\prej>\frac\gamma\alpha (1-2\eps)$ : only $\ratiodiff(\xx(t))>0$ seems possible and we should  observe only \up~trajectories.
\end{enumerate}
%
%

This reasoning  matches simulations in Figure~\ref{fig:multi_selfconsist_example} very well. 

\subsection{The impact of recommendations on opinions}
In this section we make some important observations about the effects of the recommendations on the evolution of the opinions. Statistically, we observe that recommendations produce a significant polarizing effect on the opinions and that this effect is closely entangled with their effectiveness in terms of increasing the click-through rate.

Our first observation is that {\it most trajectories produce opinions that are more extreme than the prejudices}, that is, they exhibit polarizing effect.
\begin{figure}\centering
\includegraphics[page =30, trim={\COMPfigtrimlASY} {\COMPfigtrimbASY} {\COMPfigtrimrASY} {\COMPfigtrimtASY}, clip ,width={\COMPfigwidthASY}, keepaspectratio=true]{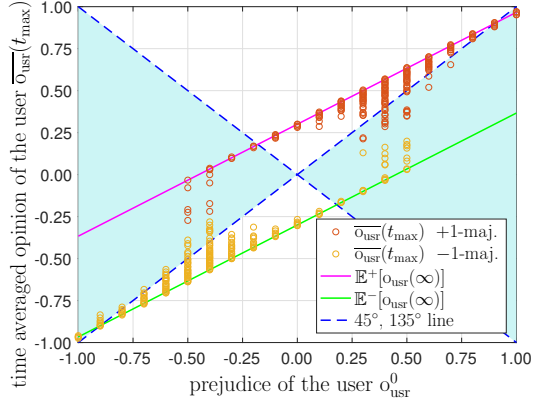}
\caption{Plot of final opinions against prejudices (initial opinions). In shaded areas, the time averaged opinion $\aveopi(\tmax)$ is {\it less extreme} than the prejudice $\prej$, i.e., $|\aveopi(\tmax)| \leq |\prej|$;  in non-shaded areas, it is {\it more extreme}. Parameters: $\alpha = 0.20$, $\beta = 0.70$, $\epsilon = 0.05$, 1000 simulations.}\label{fig:multi_milder_extremer}
\end{figure}
%
In Figure~\ref{fig:multi_milder_extremer} the shaded areas correspond to opinions that got milder during their evolution, i.e., $|\aveopi(\tmax)| \leq |\prej|$. 
It is clear that a large majority of the trajectories falls in the non-shaded areas, thereby implying that for most choices of the parameters, recommendations make opinions more extreme, that is, contribute to polarize opinions towards either $-1$ or $+1$.

\begin{figure}\centering
\includegraphics[page =31, trim={\COMPfigtrimlASY} {\COMPfigtrimbASY} {\COMPfigtrimrASY} {\COMPfigtrimtASY}, clip ,width={\COMPfigwidthASY}, keepaspectratio=true]{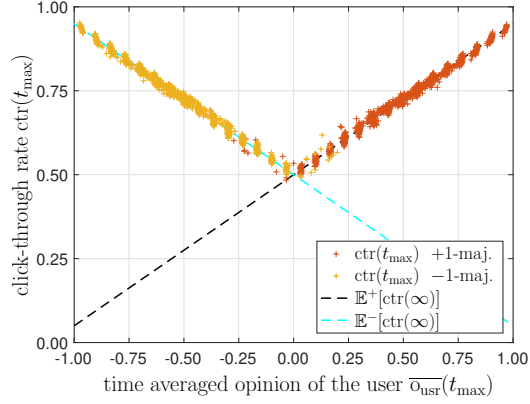}
\caption{
Click-through rate $\ctr(\tmax)$ with respect to time-averaged opinion $\aveopi(\tmax)$, with the simulations  distinguished by their \up~and \down~character. The equations of the shaded black and cyan line are \eqref{eq:Ep-click} and \eqref{eq:Em-click}, respectively.
Parameters: $\alpha = 0.20$, $\beta = 0.70$, $\epsilon = 0.05$, $\prej \in \{-1.00, -0.90, \ldots, 1.00\}$: for each prejudice $\prej$ we show 1000 realizations with $\tmax =5000$}
\label{fig:opinion-ctr}
\end{figure}

Our second observation is that 
{\it recommendations are more effective when opinions are more extreme.}
Figure~\ref{fig:opinion-ctr} represents the click-through rate $\ctr(\tmax)$ versus the time averaged opinion  $\aveopi(\tmax)$. We observe that the dashed lines in the plot, with expressions \eqref{eq:Ep-click} and \eqref{eq:Em-click}, correspond to the two possible weighted averages of \eqref{eq:g-fun-example}, i.e.
$$(1-\epsilon)\inter(\opi,\plusone)  + \epsilon\inter(\opi,-1)  \quad\text{and} \quad \epsilon\inter(\opi,\plusone)  + (1-\epsilon)\inter(\opi,-1)\,.  $$ and that the realizations concentrate on theoretical predictions in their portions above $0.50$. These simulations confirm that the recommender system increases the ratio of collected clicks and that this ability is enhanced by extreme opinions.

%



\begin{figure}\centering
\includegraphics[page =24, trim={\COMPfigtrimlASY} {\COMPfigtrimbASYX} {\COMPfigtrimrASY} {\COMPfigtrimtASY}, clip ,width={\COMPfigwidthASY}, keepaspectratio=true]{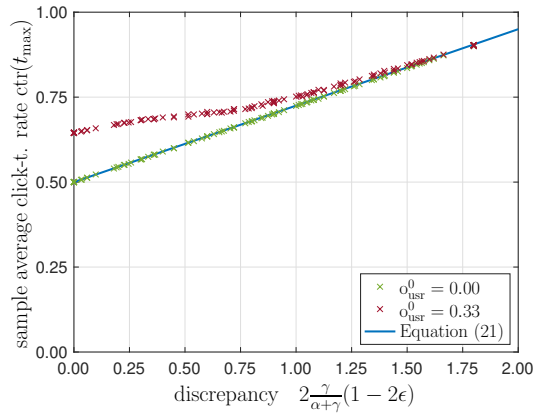}
\caption{Sample average of the click-through rate $\ctr(\tmax)$ against the discrepancy. The different values of the discrepancy, defined in Equation \eqref{eq:discrepancy}, have been obtained with 116 different combinations of the parameters $\alpha,\beta, \gamma$,  
 see Appendix \ref{sect:appendix} for details, while keeping $\epsilon = 0.05$. Each cross represents 
 1000 simulations with $\tmax =1000$. 
}
\label{fig:mean-ctr-discrep} 
\end{figure}

Our third observation sheds more light on the connection between recommender systems and polarizing effect. Not only we have already observed that recommendations are more effective when opinions are extreme, but actually we find an explicit {\em correlation between the effectiveness of recommendations and their impact on opinions}.
Indeed, the average click-through rate is such that
\begin{align}
\lim_{t \to \infty} \frac12 \left( \Ep[\ctr(t)] + \Em[\ctr(t)] \right) & =
\frac12 + \frac12 \frac{\gamma}{\alpha+\gamma}(1-2\eps)^2 \nonumber
\\ & = \frac12 + \frac14 (1-2\eps) \lim_{t \to \infty} \left( \Ep[ \opi(t)] - \Em[\opi(t)] \right) \label{eq:mean-ctr-discrep}
\end{align}
where we used \eqref{eq:discrepancy} to recognise the \textit{discrepancy}. 
This unweighted average is relevant if the prejudice $\prej$ is zero: simulations suggest that in such case the trajectories are equally likely to be \up~and \down, a fact compatible with the symmetry of $\prej = 0$. Indeed, the analysis is confirmed by simulations:
%
%
%
in Figure~\ref{fig:mean-ctr-discrep} we plot the sample average of the 
click-through rates  $\ctr(\tmax)$ against the \textit{discrepancy}. 
The simulation outcomes corresponding to the null prejudice $\prej = 0.00$ match expression \eqref{eq:mean-ctr-discrep}, 
while the simulation outcomes corresponding to prejudice $\prej = 0.33$ are distributed above the former and confirm that the {\em click-through rate is monotonically increasing with the discrepancy}.

Our fourth and final observation is that the tuning parameter $\epsilon$ has a crucial influence on both the effectiveness of the recommender systems and on its side effects, and can be used to mitigate them. Indeed,  $\epsilon$ controls the amount of randomness injected in the system. If $\epsilon=0.5$, we have random recommendations that propose $\plusone$ and $-1$ with equal probabilities. For a value of $\epsilon$ in $(0,\frac12)$, the algorithm favors the option that stimulated a larger interest in the user; this bias gets larger for the smaller $\epsilon$. According to our analysis, this recommendation bias produces an average user opinion that differs from its unbiased counterpart ($\eps=0.5$) by an {\em opinion distortion}
\begin{equation}\label{eq:deltao}\deltao:=\E^{\pm}[\opi(\infty);\epsilon]-\E^{\pm}[\opi(\infty);\epsilon=0.5]=\pm\frac{\gamma}{\alpha + \gamma} (1-2\eps), \end{equation}
i.e., half of the \textit{discrepancy}. 
At the same time, the recommendation bias produces a click-through rate larger than in the unbiased case by a {\em click-through rate gain}
\begin{equation}\label{eq:deltactr}\gammactr:=\E^{\pm}[\ctr(\infty);\epsilon]-\E^{\pm}[\ctr(\infty);\epsilon=0.5] = \pm \frac12 \frac{\alpha}{\alpha+\gamma}\prej (1-2\epsilon) + \frac12 \frac{\gamma}{\alpha+\gamma} (1-2\epsilon)^2\,.\end{equation}
Both these quantities are monotonically increasing in $(1-2\eps)$. Hence, our analysis suggests that {\em mitigating the impact  of the recommender system on the opinions has a price in terms of the achievable click-through rate}.
To make this connection clearer, we can combine \eqref{eq:deltao} and \eqref{eq:deltactr} and deduce
\begin{equation}\label{eq:delta-relation}\gammactr= \frac12 \frac{\alpha}{\gamma}\prej \deltao+ \frac12 \frac{\alpha+\gamma}{\gamma} (\deltao)^2\,,\end{equation}
thereby showing that the relation between the the opinion distortion $\deltao$ and the click-through rate gain $\gammactr$ is independent of $\eps$ and therefore depends only on the characteristics of the user. This relation is well matched by the simulations in Figure~\ref{fig:deltas}.

\begin{figure}\centering
\includegraphics[page =37, trim={\COMPfigtrimlASY} {\COMPfigtrimbASYX} {\COMPfigtrimrASY} {\COMPfigtrimtASY}, clip ,width={\COMPfigwidthASY}, keepaspectratio=true]{compiled_figures.pdf}
\caption{Empirical click-through rate gain against empirical opinion distortion, with the simulations  distinguished by their \up~and \down~character. 
The empirical ``ctr gain'' is $[\ctr(\tmax);\epsilon]$ minus the sample average of $[\ctr(\tmax);\epsilon=0.5]$. 
The empirical ``op. distortion'' is $[\aveopi(\tmax);\epsilon]$ minus the sample average of $[\aveopi(\tmax);\epsilon=0.5]$. 
In the simulations, $\alpha=0.20$, $\beta=0.70$, $\gamma=0.10$, $\prej= 0.33$ and $\tmax=5000$. There are 100 simulations for each  values of $\eps$ in the set $\{0.001, 0.0025, 0.005, 0.0075, 0.01, 0.025, 0.05, 0.075, 0.1, 0.15, 0.2, 0.25, 0.3, 0.35, 0.4, 0.45, 0.5\}$. The aforementioned sample averages are computed over all simulations with $\eps = 0.5$.
}
\label{fig:deltas}
\end{figure}

\section{Conclusion}\label{sect:outro}

In this paper, we have proposed a mathematical model of the interaction between a user and an online service providing personalized recommendations. The model is simple enough to allow for its analytical treatment, while being rich enough to include the main features of the user-platform interaction, which we identified on the user side as information assimilation and confirmation bias, and on the recommender side as measuring the user's engament and learning through exploration-exploitation. Our results quantify the effects of personalization on the evolution of user's opinion, showing that the personalized recommendations typically drive the user towards more extreme opinions. 

While we believe that our model is informative enough to make it relevant in the heating debate on the impact of machine learning on our societies, we are well aware of its limitations. Indeed, our model describes the behavior of a single user, but real recommender systems deal with large numbers of users that can have social ties and shared interests.
Our focus on a single user has allowed us to highlight the feedback loop between the user's opinion and the recommendations, but has limited the scope of our work in two ways. 
First, our recommender system was not allowed to exploit either social ties or shared interests to provide its recommendations. Instead, real recommender systems are {\em collaborative} and effectively take advantage of these features~\cite{JZFF:2010:book-recommender}: this fact has been included in some mathematical models \cite{DGL:2013:goel,BLPRT:2016:limit-recommendations}.
Second, recommendations were the only drive to the opinion dynamics in our model. Instead, opinion dynamics are also driven by social interactions (both directly and through the collaborative elements of the recommender system), creating a complex entanglement of effects. On this matter, we note that some experimental studies on Facebook have reported that ideological contents are primarily filtered by user's social connections rather than by the curation algorithms, suggesting that user preferences may have stronger impact than algorithmic personalisation \cite{Bakshy:2015:exposure}. A future model that includes both social and recommendation effects could shed more light on this issue. 

%

\appendix
\section{Dependence on opinion model's parameters} \label{sect:appendix}
\begin{figure}[!t]
\centering
\boxfig{\includegraphics[trim={0.9\figtriml} {0.9\figtrimb} {0.9\figtrimr} {0.9\figtrimt},
	clip,width={\figwidth}, keepaspectratio=true]{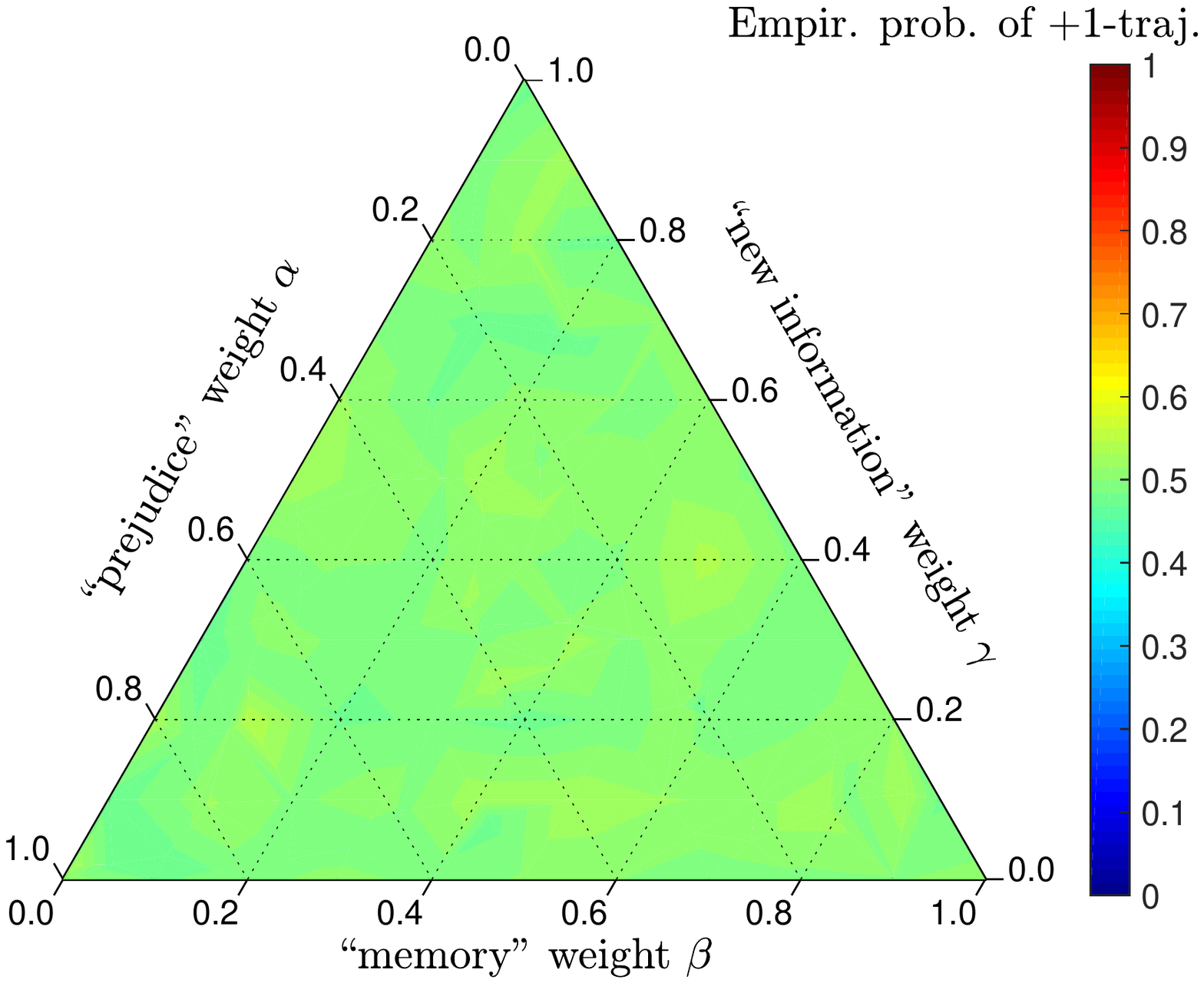}} 
\boxfig{\includegraphics[trim={0.9\figtriml} {0.9\figtrimb} {0.9\figtrimr} {0.9\figtrimt},
	clip,width={\figwidth}, keepaspectratio=true]{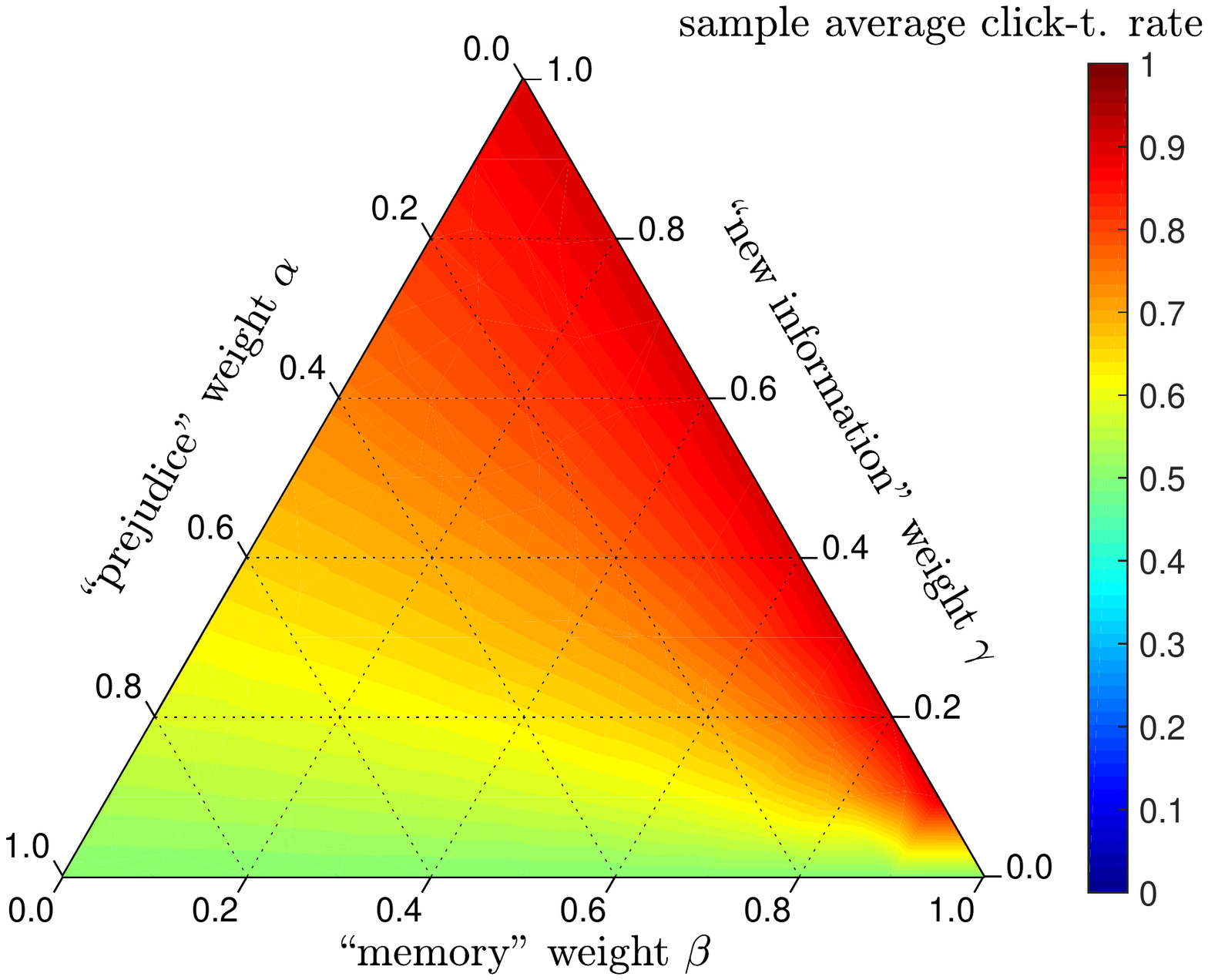}} 
	\caption{\label{fig:tri_z0_00}
	Simulations with prejudice $\prej = 0.00$ and various combinations of the parameters $\alpha,\beta, \gamma$ covering the 2-simplex. 
	To each combination correspond 1000 simulations up to $\tmax =1000$, with position $\pos(t)$ recommended using $\epsilon = 0.05$.
	Left: the empirical probability of \up~trajectory, that is about 0.50. 
	Right: the sample average of the click-through rate values $\ctr(\tmax)$.
}
\end{figure}

\begin{figure}[!t]
\centering
\boxfig{\includegraphics[trim={0.9\figtriml} {0.9\figtrimb} {0.9\figtrimr} {0.9\figtrimt},
	clip,width={\figwidth}, keepaspectratio=true]{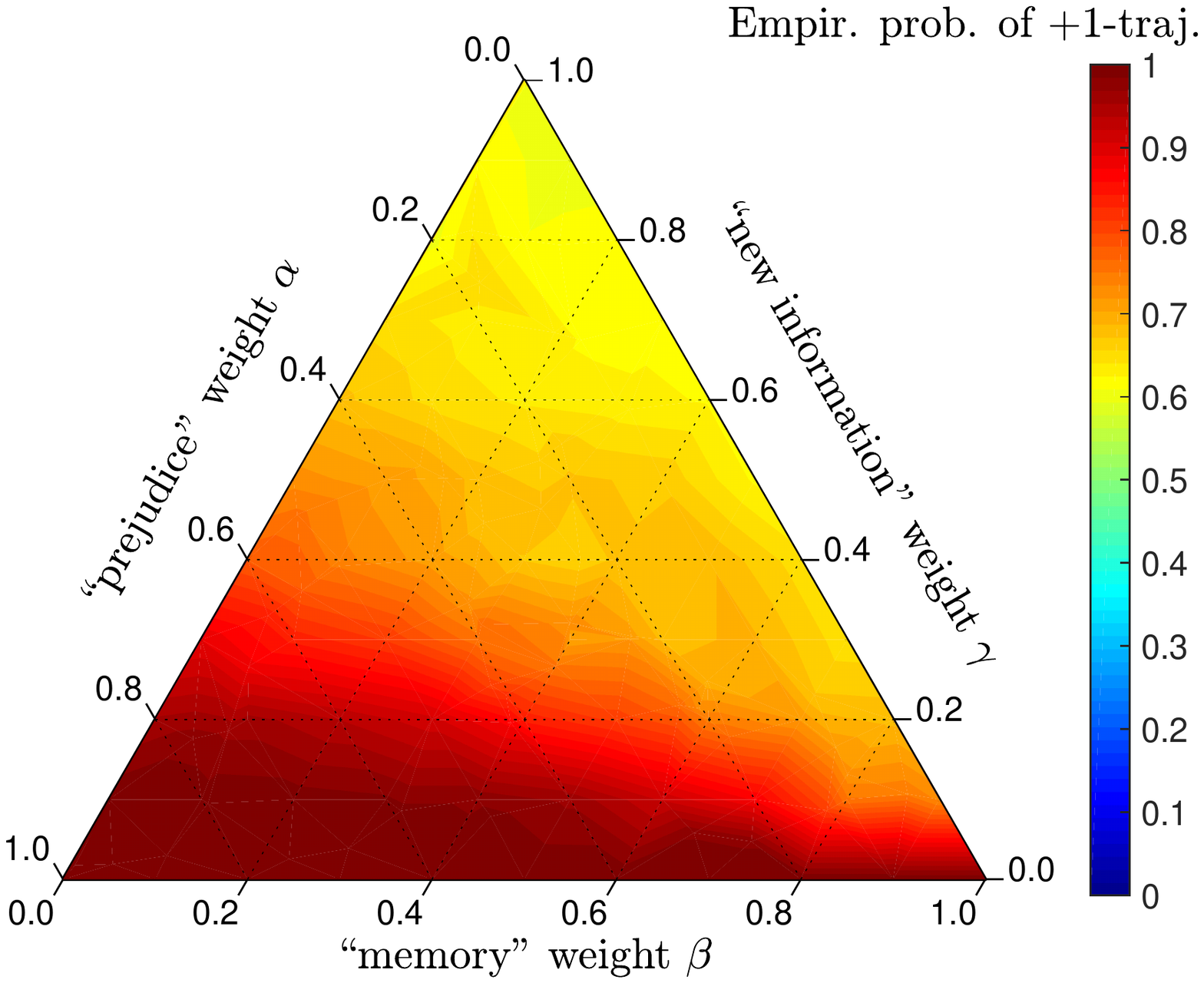}} 
\boxfig{\includegraphics[trim={0.9\figtriml} {0.9\figtrimb} {0.9\figtrimr} {0.9\figtrimt},
	clip,width={\figwidth}, keepaspectratio=true]{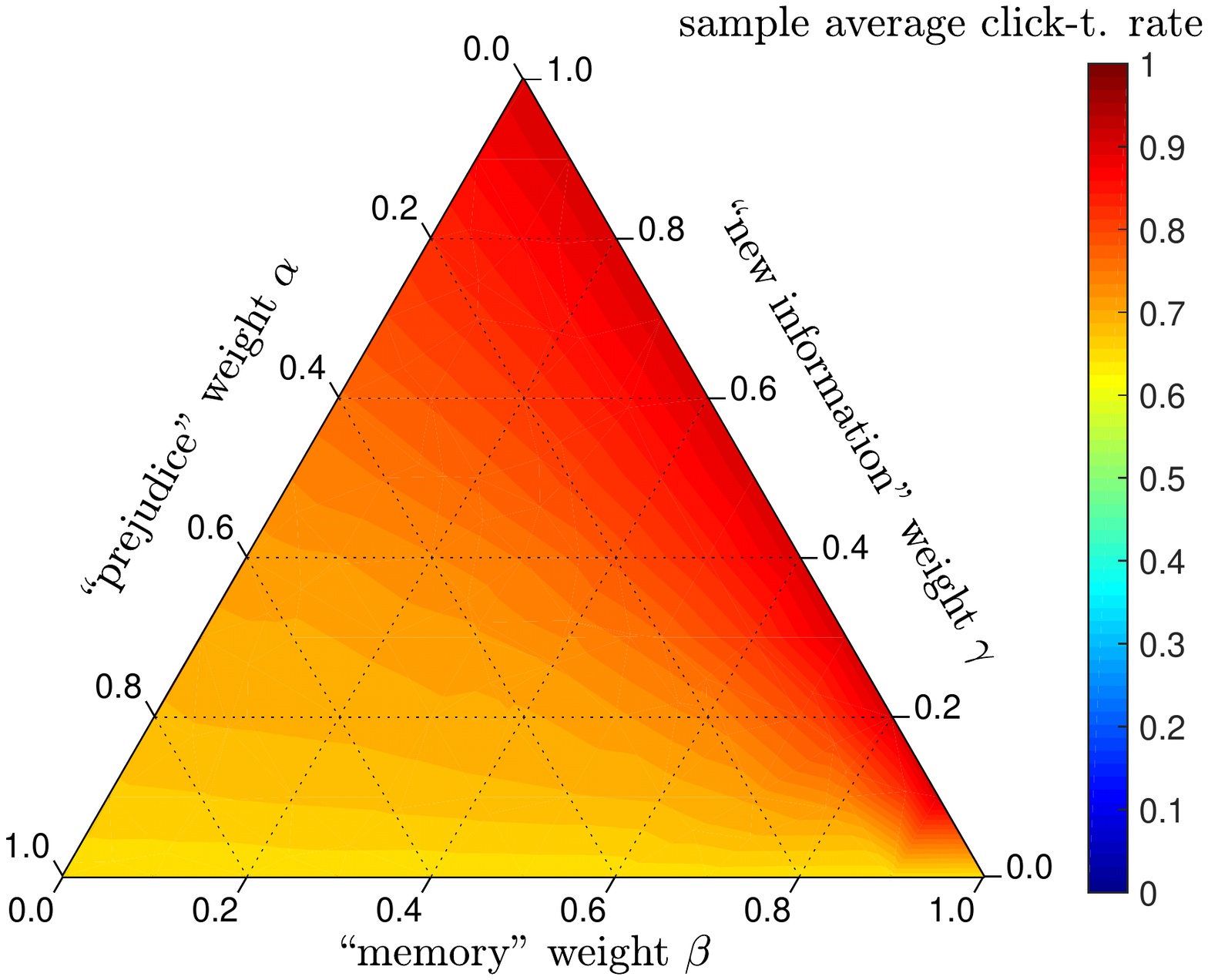}} 
\caption{\label{fig:tri_z0_33}
	Simulations with prejudice $\prej = 0.33$ and various combinations of the parameters $\alpha,\beta, \gamma$ covering the 2-simplex. 
	To each combination correspond 1000 simulations up to $\tmax =1000$, with position $\pos(t)$ recommended using $\epsilon = 0.05$.
	Left: the empirical probability of \up~trajectory. 
	Right: the sample average of the click-through rate values $\ctr(\tmax)$.
}
\end{figure}

Along the paper, we have made a running choice of $\alpha$, $\beta $ and $\gamma$, but it is clear that different users can be characterised by different parameters. 
We therefore explore with simulations the dependence of the results on these parameters.
In such exploration we keep $\epsilon = 0.05$ and take two non-negative values of the prejudice, i.e., $\prej =0.00$ (see Figure~\ref{fig:tri_z0_00}) and $0.33$ (see Figure~\ref{fig:tri_z0_33}): for each value we repeat the simulations with different $\alpha$, $\beta $ and $\gamma$. %
For the simulations we pick 116 points on the 2-simplex $\{(\alpha, \beta, \gamma) : \alpha,\beta,\gamma \geq 0; \alpha+\beta+\gamma=1\}$: 66 points lie on the grid with spacing 0.10 while 50 points are randomly chosen. 
For each combination of parameter we run 1000 simulations up to $\tmax = 1000$ and compute two variables of interest, namely the empirical probability of obtaining a \up~trajectory and 
the sample average of the click-through rates $\ctr(\tmax)$. 
In the following, we use triangular colormap plots to represent these two variables of interest with respect to $\alpha, \beta, \gamma$.
The triangles represent the 2-simplex and 
the colours encode the values of the variables of interest.

Figure~\ref{fig:tri_z0_00} correspond to $\prej = 0.00$. 
The empirical probability of obtaining a \up~trajectory is about 0.5 for all combinations of $\alpha$, $\beta $ and $\gamma$, a value compatible with the null prejudice for symmetry reasons.
In a similar way, the sample average of the  click-through rate increases gradually from 0.5 on the edge $\gamma = 0$  to 0.9 on the edge $\alpha=0$. We recall that on the edge $\gamma=0$ the new information $\pos(t)$ has no role in the opinion model \eqref{eq:opinion-model}, while on the edge $\alpha=0$ the prejudice $\prej$ has no influence beside being the initial opinion. 
Figure~\ref{fig:tri_z0_33} correspond to $\prej = 0.33$. 
The empirical probability of obtaining a \up~trajectory takes on values between 0.5 and 1; it decreases with $\gamma - \alpha$ and shows a straight boundary with a sharp transition. 
The trends of 
 the sample average of the click-through rate with respect to the parameters $\alpha, \beta, \gamma$ is again increasing from the edge $\gamma=0$ to the edge $\alpha=0$. 
If $\gamma$ is small and $\alpha$ large, the \up~probability is one. With such parameters, the opinion $\opi(t)$ remains in any case similar to $\prej$, which is positive: this favours the position $\plusone$, making the \up~trajectory the unique possibility. 
We have also performed simulations for other values of $\prej$, obtaining results that are qualitatively similar to those we observe for $\prej =0.33$ (clearly, negative values of $\prej$ produce results that can be deduced by symmetry from positive ones). 

Our exploration of the parameter space confirms the validity of the simulations shown in the rest of the paper and indicates that the values of the parameters influence the evolution in a rather intuitive way: namely, larger weights on the prejudice make easier for the recommender system to identify whether $+1$ or $-1$ is the best recommendation, but reduce the recommender's effectiveness in reaping clicks.

%
%

%

\end{document}